\documentclass[12pt,preprint]{aastex}

\def\simlt{\lower.5ex\hbox{$\; \buildrel < \over \sim \;$}}
\def\simgt{\lower.5ex\hbox{$\; \buildrel > \over \sim \;$}}

\def\hnot{\ifmmode H_0 \else H$_0$ \fi}

\def\msun{\ifmmode {\rm M_\odot} \else M$_\odot$\fi}
\def\lsun{\ifmmode {\rm L_\odot} \else L$_\odot$\fi}

\def\deg{\ifmmode ^{\circ}
         \else $^{\circ}$\fi}
\def\pdeg{\ifmmode
           $\setbox0=\hbox{$^{\circ}$}\rlap{\hskip.11\wd0 .}$^{\circ}
     \else \setbox0=\hbox{$^{\circ}$}\rlap{\hskip.11\wd0 .}$^{\circ}$\fi}

\def\msunyr{\ifmmode {\rm M_\odot~yr^{-1}}\else${\rm M_\odot~yr^{-1}}$\fi}
\def\lam{\ifmmode {\lambda} \else {$\lambda$} \fi}
\def\lamLlam{\ifmmode \lambda L_{\lambda}(5100) \else {$\lambda L_{\lambda}(5100)$} \fi}
\def\nuLnu{\ifmmode \nu L_{\nu}(5100) \else {$\nu L_{\nu}(5100)$} \fi}

\def\mdoto{\ifmmode {\dot{M}_0} \else  $\dot{M}_0$ \fi}
\def\teff{\ifmmode {T_{eff}} \else $T_{eff}$ \fi}
\def\ilam{\ifmmode {I_\lambda} \else  $I_\lambda$ \fi}
\def\inu{\ifmmode {I_\nu} \else  $I_\nu$ \fi}
\def\fnu{\ifmmode {F_\nu} \else  $F_\nu$ \fi}

\def\yr{\ifmmode {\rm yr} \else  yr \fi}
\def\cm{\ifmmode {\rm cm} \else  cm \fi}
\def\cmmitwo{\ifmmode \rm cm^{-2} \else $\rm cm^{-2}$\fi}
\def\cmmithree{\ifmmode \rm cm^{-3} \else $\rm cm^{-3}$\fi}
\def\cmps{\ifmmode \rm cm~s^{-1}\else $\rm cm~s^{-1}$\fi}
\def\cmpsps{\ifmmode \rm cm~s^{-2}\else $\rm cm~s^{-2}$\fi}
\def\kmps{\ifmmode \rm km~s^{-1}\else $\rm km~s^{-1}$\fi}
\def\kmpspmpc{\ifmmode \rm km~s^{-1}~Mpc^{-1} \else
    $\rm km~s^{-1}~Mpc^{-1}$\fi}
\def\ergps{\ifmmode \rm erg~s^{-1} \else $\rm erg~s^{-1}$ \fi}
\def\ergpspcm{\ifmmode \rm erg~s^{-1}~cm^{-2} \else $\rm erg~s^{-1}~cm^{-2}$ \fi}
\def\ergpspcmphz{\ifmmode \rm erg~s^{-1}~cm^{-2}~Hz^{-1} \else $\rm
erg~s^{-1}~cm^{-2}~Hz^{-1}$ \fi}
\def\ergpspcmpa{\ifmmode \rm erg~s^{-1}~cm^{-2}~\AA^{-1} \else $\rm
erg~s^{-1}~cm^{-2}~\AA^{-1}$ \fi}
\def\ergpsphz{\ifmmode \rm erg s^{-1} Hz^{-1} \else
   $\rm erg s^{-1} Hz^{-1}$ \fi}

\def\mbh{\ifmmode M_{\mathrm{BH}} \else $M_{\mathrm{BH}}$\fi}
\def\mbulge{\ifmmode M_{\mathrm{bulge}} \else $M_{\mathrm{bulge}}$\fi}

\def\mbhsigstar{\ifmmode M_{\mathrm{BH}} - \sigma_* \else $M_{\mathrm{BH}} - \sigma_*$ \fi}
\def\deltalogmbh{\ifmmode \Delta {\mathrm{log}} M_{\mathrm{BH}} \else $\Delta$ log $M_{\mathrm{BH}}$ \fi}

\def\sigstar{\ifmmode \sigma_* \else $\sigma_*$\fi}
\def\sigthree{\ifmmode \sigma_{\mathrm{[O~III]}} \else $\sigma_{\mathrm{[O~III]}}$\fi}
\def\sigtwo{\ifmmode \sigma_{\mathrm{[O~II]}} \else $\sigma_{\mathrm{[O~II]}}$\fi}
\def\sigco{\ifmmode \sigma_{\mathrm{CO}} \else $\sigma_{\mathrm{CO}}$\fi}
\def\sigstarco{\ifmmode \sigma_{\mathrm{*CO}} \else $\sigma_{\mathrm{*CO}}$\fi}

\def\fwzi{\ifmmode {FWZI} \else $FWZI$ \fi}
\def\fwhm{\ifmmode {FWHM} \else $FWHM$ \fi}
\def\whbeta{\ifmmode {FWHM(\mathrm{H}beta)} \else $FWHM(\mathrm{H}\beta)$ \fi}
\def\wco{\ifmmode {FWHM(\mathrm{CO})} \else $FWHM(\mathrm{CO})$ \fi}
\def\wthree{\ifmmode {FWHM(\mathrm{[O~III]})} \else $FWHM(\mathrm{[O~III]})$ \fi}
\def\wtwo{\ifmmode {FWHM(\mathrm{[O~II]})} \else $FWHM(\mathrm{[O~II]})$ \fi}
\def\mstarco{\ifmmode M_{\mathrm *CO} \else $M_{\mathrm *CO}$ \fi}
\def\mstarsig{\ifmmode M_\sigma \else $M_\sigma$ \fi}
\def\mthree{\ifmmode M_{\mathrm [O III]} \else $M_{\mathrm [O III]}$ \fi}
\def\mtwo{\ifmmode M_{\mathrm [O II]} \else $M_{\mathrm [O II]}$ \fi}
\def\hbeta{\ifmmode {\rm H}\beta \else H$\beta$\fi}

\def\log{\rm{log}}

\newcommand{\civ}{C~{\sc iv}}
\newcommand{\oiii}{[O~{\sc iii}]}

\newcommand{\mgii}{Mg~{\sc ii}}

\slugcomment{Submitted to The Astrophysical Journal}

\shorttitle{QSOs at High Redshift}
\shortauthors{Shields et al.}

\begin{document}

\title{The Black Hole - Bulge Relationship
                for QSOs at High Redshift}

\author{
G. A. Shields\altaffilmark{1},
K. L. Menezes\altaffilmark{1},
C. A. Massart\altaffilmark{1},
P. Vanden Bout\altaffilmark{2}
}

\altaffiltext{1}{Department of Astronomy, University of Texas, Austin, TX 78712; shields@astro.as.utexas.edu}

\altaffiltext{2}{National Radio Astronomy Observatory, 520 Edgemont Road, 
Charlottesville VA 22903; pvandenb@nrao.edu}

\begin{abstract}
We examine the black hole mass - galaxy bulge relationship in high-redshift QSOs.  Black hole masses are derived from the broad emission lines, and the host galaxy stellar velocity dispersion \sigstar\ is estimated from the widths of the radio CO emission lines.  At redshifts $z > 3$, the CO line widths are narrower than expected for the black hole mass, indicating that these giant black holes reside in undersized bulges by an order of magnitude or more.  The largest black holes ($\mbh > 10^9$~\msun) evidently grow rapidly in the early universe without commensurate growth of their host galaxies.  CO linewidths offer a unique opportunity to study AGN host galaxy dynamics at high redshift.

\end{abstract}


\section{Introduction}

The evolution of galaxies and their central supermassive black holes (SMBH) is a major topic of current interest (see review by Combes 2005).  The tight correlation of the black hole mass \mbh\ and the luminosity and the velocity dispersion \sigstar\ of the host galaxy's bulge (Gebhardt et al. 2000a; Ferrarese \& Merritt 2000; review by Kormendy \& Gebhardt 2001) points to a close evolutionary relationship.   This has inspired a number of theoretical investigations involving growth of the black hole by accretion of gas. When the luminosity of the active galactic nucleus (AGN) becomes large enough, it drives the remaining gas from the nucleus of the host galaxy, ending black hole growth as well as star formation
(Silk \& Rees 1998;  Di Matteo et al. 2005, and references therein).  Numerical simulations by Di Matteo et al. find a tight \mbhsigstar\ relationship from simple assumptions about heating by the AGN luminosity leading to evacuation of residual gas.

The \mbhsigstar\ relationship for nearby galaxies  is given by Tremaine et al. (2002) as
\begin{equation}
\label{e:tremaine}
\mbh = (10^{8.13}~\msun)(\sigstar/200~\kmps)^{4.02},
\end{equation}
with black holes spanning the range roughly $10^5-10^9$ \msun.  One clue to the origin of this relationship is its evolution over cosmic time.  
Shields et al. (2003, hereinafter S03) investigated the \mbhsigstar\ relationship 
in QSOs at high redshift by estimating \mbh\ and \sigstar\ from the \hbeta\ and \oiii\ emission lines, respectively.  They found that the local \mbhsigstar\ relationship is obeyed, in the mean, up to masses approaching $10^{10}~\msun$ and at redshifts up to $z = 3.3$. Salviander et al.  (2005) similarly find little evolution in the \mbhsigstar\ relationship in QSOs at redshifts up to $z = 1.1$.    However, the observation of luminous quasars up to redshifts $z = 6.4$ (Fan et al. 2001) shows that large black holes ($\sim 10^9~\msun$) can grow rapidly in the early universe (Haiman \& Loeb 2001; Volonteri \& Rees 2005).   This raises the question of whether, even at such early times, these massive black holes reside in commensurate galaxies so as to obey the local \mbhsigstar\ relationship.  The use of \oiii\ as a surrogate
for \sigstar\ (see below) has a practical limit of $z \approx 3$, as the \oiii\ lines are shifted beyond the infrared K-band window for higher redshifts.  However, the radio CO emission lines have been observed in a number of high redshift QSOs and radio galaxies
(Solomon \& Vanden Bout 2005, hereinafter SV05).  To the extent that the CO line widths reflect orbital motion in the gravitational potential of the host galaxy, this affords an opportunity to assess the \mbhsigstar\ relationship at a cosmic time of only one billion years.

In this paper, we investigate the relationship between \mbh\ and CO linewidth for high redshift QSOs as a means of assessing the \mbhsigstar\ relationship at early times.  All values of luminosity used in this study are calculated using the cosmological parameters $\hnot = 70~\kmpspmpc, \Omega_{\rm M} = 0.3$, and $\Omega_{\Lambda} = 0.7$.

\section{QSO Black Hole Masses and Velocity Dispersions}

\subsection{Black Hole Masses}\label{s:mbh}

Derivation of SMBH masses from AGN broad emission-line widths is now well established (see S03; Kaspi et al. 2005,  and references therein).    The broad line widths are assumed to result from orbital motion around the black hole, so that  $\mbh = v^2R/G$ with $v = f \times\,\fwhm$.   The factor $f$ depends on the geometry and kinematics of the broad line region  (BLR) (McLure \& Dunlop 2001), with $f = \sqrt3/2$ for random orbits.  The radius of the broad line region (BLR) is estimated on the basis of echo-mapping studies that give
$R \propto L^{\gamma}$ with $\gamma = 0.5 - 0.7$ 
(Kaspi et al. 2000), where $L$ is the continuum luminosity.   For ease of comparison with S03, we adopted their formula
\begin{equation}
\label{e:mbh}
\mbh = (10^{7.69}~\msun)v_{3000}^2 L_{44}^{0.5},
\end{equation}
where $v_{3000} \equiv$ \whbeta/3000 \kmps\ and the BLR continuum luminosity at 5100~\AA\ is $L_{44} \equiv \nuLnu /(10^{44}~\ergps)$.  A larger value of $\gamma$ would increase \mbh\ in these luminous QSOs and
aggravate the discrepancy that we find between \mbh\ and CO line width.
 Black hole masses derived in this way have been used in a variety of studies of QSOs and black hole demographics (e.g., Vestergaard 2002; McLure \& Dunlop 2004).

 Line widths of \mgii\ and \hbeta\ agree sufficiently well to allow \mbh\ determinations from \mgii\ (McLure \& Jarvis 2002; McClure \& Dunlop 2004; Salviander et al. 2005).   Vestergaard (2002) finds that \civ\ also is a useful alternative to \hbeta\ for measurements of \mbh.  Use of these lines allows black hole mass determinations at higher redshift.

\subsection{Velocity Dispersion Estimated from \oiii}\label{s:sigma}

For QSOs, \sigstar\ is difficult to measure because of the faintness of the host galaxy and the glare of the active nucleus.  Alternative means of assessing \sigstar\ are therefore needed.  Nelson \& Whittle (1996) and Nelson (2000) find that, with considerable scatter, the width of the \oiii$\lambda5007$ lline, 
$\sigthree \equiv {\mathrm \wthree/2.35}$,
tracks \sigstar.  Bonning et al. (2005) show that in the mean, \sigthree\ agrees with the value of \sigstar\ expected from the host galaxy luminosity of PG QSOs, the 1$\sigma$ dispersion being 0.13~dex.

\subsection{CO Line Widths as a Surrogate for \sigstar\ in QSOs}

Motivated by the utility of \sigthree, we define 
$\sigco \equiv \wco/2.35$ and ask whether it is
a useful measure of host galaxies of QSOs.
CO observations of high redshift galaxies and AGN are reviewed by SV05.
The radius of the CO emission in the high redshift QSOs analyzed below is typically one or a few kpc and is likely dominated by the bulge gravity. 
However, the CO emitting gas in QSOs is often assumed to be in a rotating disk, raising the question of inclination.  The CO line profiles of the QSOs considered here typically have $\fwhm/\fwzi \approx 2/3$, where \fwzi is the full width at zero intensity.  For an isothermal sphere, taken as an approximation to the bulge mass distribution, the circular velocity is
$v_c = \sqrt{2} \sigstar$.  If we take $\fwzi = (3/2) \fwhm = 2 v_c$, then $\sigstar = 0.53\,\fwhm$.  For random
orientations, $\mathrm{sin}\, i$ averages 0.866, giving
 $\sigstar = 0.61\fwhm$.  This is a factor 1.25 larger than \sigco\ as defined above.    This could be altered by geometrical biases, such as an obscuring torus, favoring discovery of face-on disks in optical surveys.
 
\begin{figure}[!t]
   \includegraphics[width=\columnwidth]{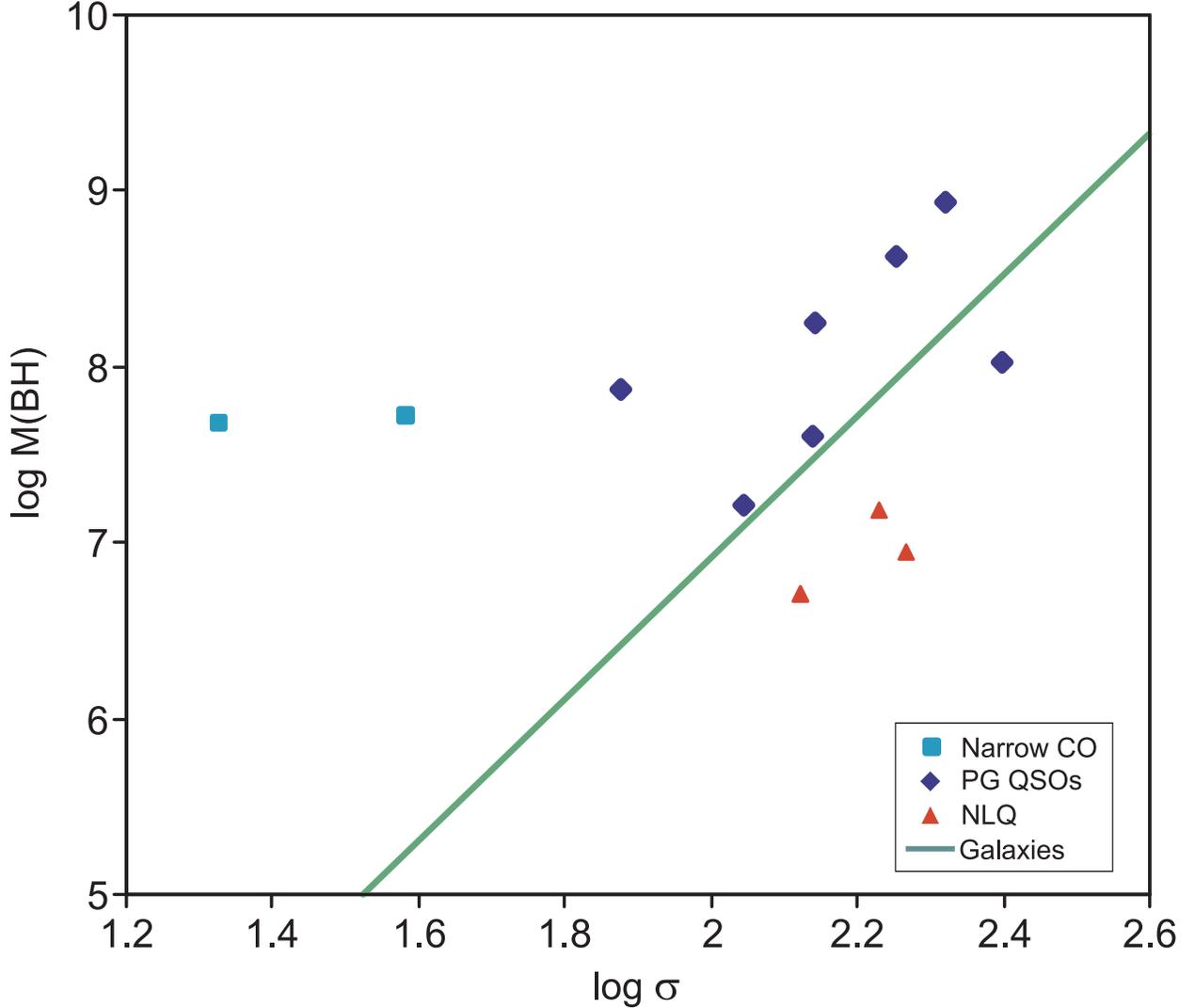}
   \caption{
The $M_{\mathrm{BH}}-\sigma_*$ relationship using
$\sigma_{\mathrm{CO}}$ as a
surrogate for $\sigma_*$ (see Table~1).  The solid line is equation \ref{e:tremaine}.
Triangles represent objects with broad line widths less than 1500~\kmps (NLQ1s).  Diamonds represent other PG QSOs.  Squares are two objects with exceptionally narrow  CO lines (see text).}
   \label{f:pg_msig}
\end{figure}

We have attempted an empirical calibration by comparing \mbh\ with a surrogate for \sigstar\ derived from the CO line widths, $\sigco \equiv \wco/2.35$ in PG QSOs. We used published CO observations  by Evans et al. (2001) and Scoville et al. (2003).  For the objects with useful CO widths, we measured \fwhm\ for \hbeta\ and \oiii\ and the continuum flux from the spectra of Marziani et al. (2003), kindly made available by P. Marziani.  Black hole masses were derived from equation (\ref{e:mbh}).  The results are given in Table~1.  Two objects objects stand apart as having very narrow CO lines,
\fwhm\ of 90 and 50 \kmps.  These ``outliers'' might be suspected of being starbursts confined to a small postion of the host galaxy, although their CO luminosities are similar to the other objects.   We present our analysis with and without these outliers.  Table 1 includes 3 QSOs with \whbeta $<$ 1500~\kmps, corresponding to the so-called narrow line Seyfert 1 galaxies (NLS1).  NLS1 are often considered a special class of AGN, and Grupe \& Mathur (2004) find that NLS1 as a group fall below the Tremaine line in Figure \ref{e:tremaine}. 
We shall call the corresponding quasars ``narrow line QSO 1s"  (NLQ1).  None of the high redshift QSOs in Table 1 has a broad line \fwhm\ less than 1500~\kmps. 

\begin{figure}[!t]
   \includegraphics[width=\columnwidth]{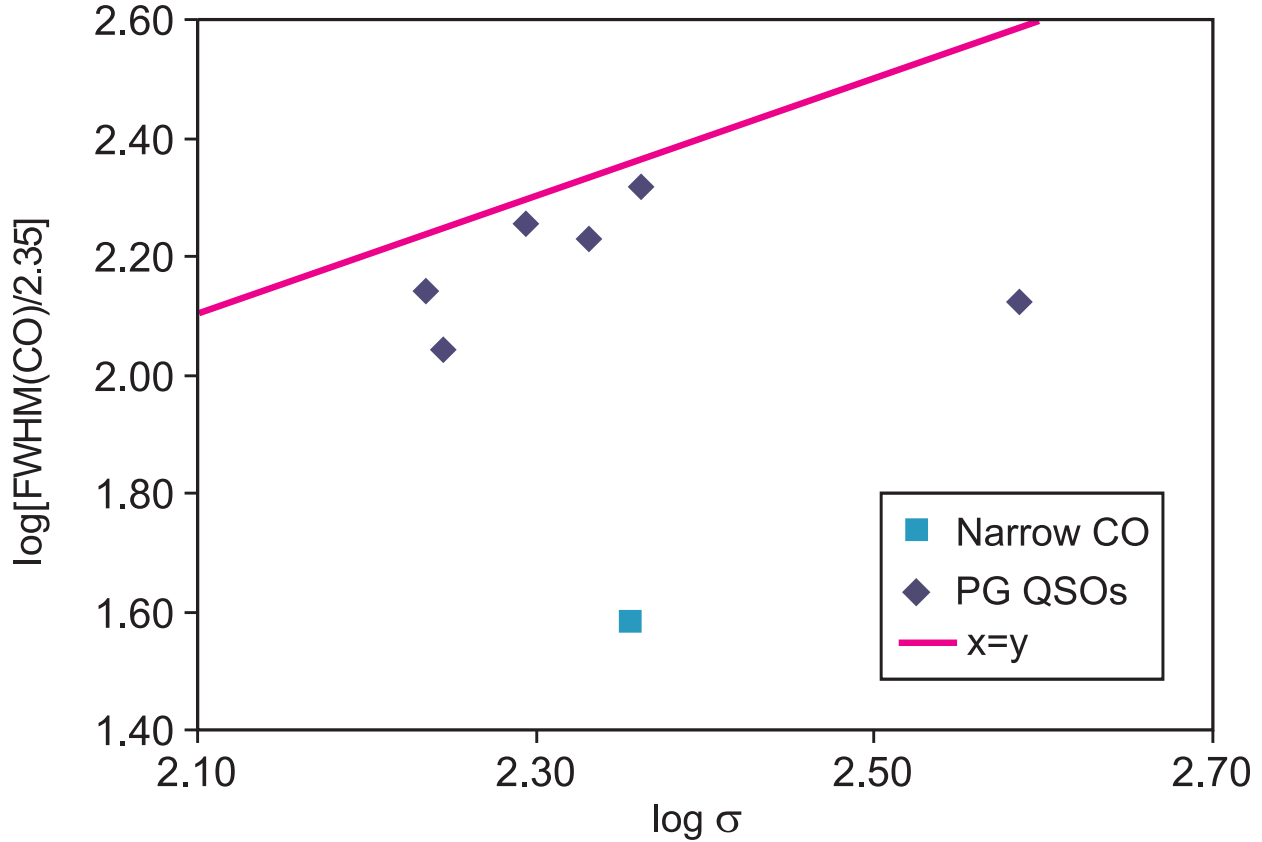}
   \caption{
  Comparison of
$\sigma_{\mathrm{[O~III]}}$ and  $\sigma_{\mathrm{CO}}$ for PG QSOs
(see text and Table~1).  The line is $\sigma_{\mathrm{[O~III]}}$ =  $\sigma_{\mathrm{CO}}$.
}
   \label{f:pg_co_oiii}
\end{figure}

We approached the calibration in three ways.  
(1)  Figure \ref{f:pg_msig} shows \mbh\ versus \sigco\ for the objects of Table~1, distinguishing the CO outliers and the NLQ1 and showing equation \ref{e:tremaine}.  We follow S03 in defining
$\deltalogmbh \equiv \log\,\mbh - \log\,\mstarsig$, where 
\mstarsig is the mass predicted by equation \ref{e:tremaine} with \sigco\
(or other surrogate) for \sigstar. We find  a mean \deltalogmbh\ of 0.57, including the outliers and the NLQ1s.  (This becomes 0.09, 1.03, or 0.48 excluding the outliers, the NLQ1s, or both, respectively.)  Assuming that the PG QSOs agree closely on average with the local \mbhsigstar\ relationship (SO3; Bonning et al. 2005), this suggests, with considerable uncertainty, that \sigco\ underestimates \sigstar\ by $\sim0.15$~dex for PG QSOs, taking account of the exponent in equation \ref{e:tremaine}.
(2) Figure \ref{f:pg_co_oiii}  compares \sigco\ with \sigthree, the latter being a statistical indicator of \sigstar\ (see above).  The
mean value of the log(\sigthree/\sigco) is 0.24, including all objects with measureable [O~III].  
(3)  Following Bonning et al. (2005), we may use the QSO host galaxy luminosity, which predicts
\sigstar\ through the Faber-Jackson relation.  We found  {\em HST} measurements of absolute magnitude for  two of the QSOs in Table~1 (see Table~1 of Bonning et al).  For these, the average inferred \sigstar\ exceeds \sigco\ by 0.14~dex.  Overall, it appears that \sigco\ should be increased by about 0.15 dex to give \sigstar,
which has the effect of reducing \deltalogmbh\ by 0.6.   However, we caution that there
could be stronger selection effects favoring face-on disks in the high redshift QSOs discussed below.  In view of the uncertainties, we will not apply any correction to \sigco; but this issue should be kept in mind.

\section{Results for High Redshift QSOs}

CO line emission from high redshift galaxies and QSOs is an important tool for studying the mass of molecular gas and its dynamics (SV05).  Some of these objects have billions of solar masses of molecular hydrogen and star formation rates up to $\sim10^3~ \msunyr$.  These objects appear to involve major mergers of gas rich galaxies that fuel the starbursts and nuclear activity, much of it hidden by dust, at least in the early stages of the outburst (Hopkins et al. 2005, and references therein).  SV05 tabulate measurements of 36 ``early universe molecular emission-line galaxies'' (EMGs), including 15 QSOs with measured CO line widths.  These have redshifts ranging from $z = 1.41$ to 6.42.  Most are gravitationally lensed objects, for which the magnification aids flux and and size measurements at the cost of some uncertainty in the magnification.  We found published optical spectra allowing useful measurements of the \fwhm\ for the \civ~$\lambda1549$ \ or \mgii~$\lambda2800$  broad emission-line for 9 of these QSOs.  We measured continuum fluxes from the optical spectra at $\lambda1350$ or $\lambda1450$ rest wavelength, and converted to \lamLlam\ with allowance for the magnification factor quoted by SV05 and an assumed powerlaw $\fnu \propto \nu^{-0.5}$ (Vanden Berk et al. 2001).  Black hole masses were then derived using equation \ref{e:mbh}.  The results are given in Table~2.

\begin{figure}[!t]
   \includegraphics[width=\columnwidth]{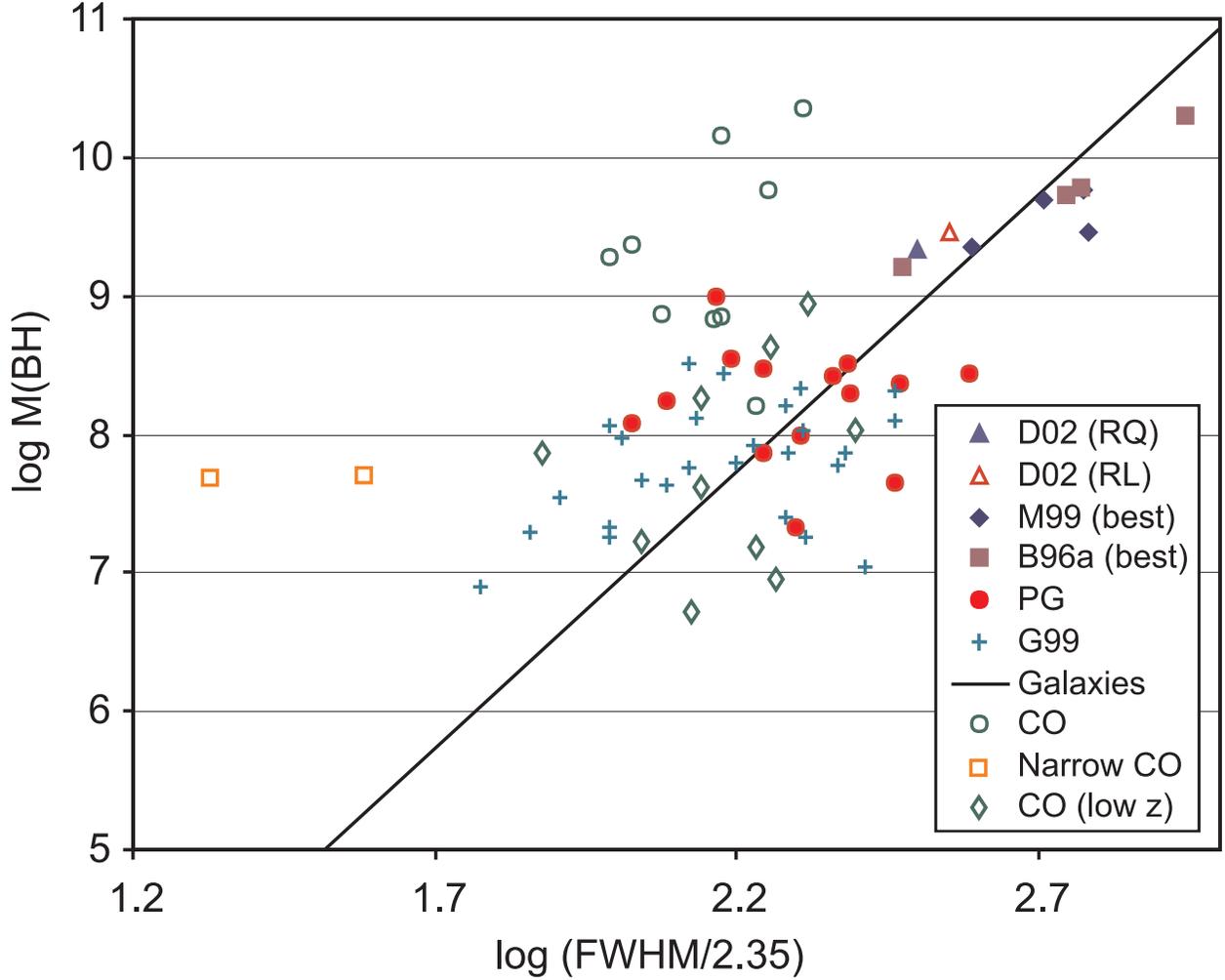}
   \caption{The $M_{\mathrm{BH}}-\sigma_*$ relationship using
$\sigma_{\mathrm{[O~III]}}$ and  $\sigma_{\mathrm{CO}}$ as
surrogates for $\sigma_*$.  Open circles are the nine
CO QSOs from this paper; open squares and diamonds are the low redshift CO QSOs.  Other points are from
Figure 2 of SO3, based on \mbh(\hbeta) and \sigthree\ (see SO3 for legend and
references).  The solid line is equation \ref{e:tremaine}.}
   \label{f:co_msig}
\end{figure}

Figure \ref{f:co_msig} shows the derived values of \mbh\ plotted against 
$\sigco$ along with the results of S03.  Most of the CO objects show \sigco\ considerably smaller than expected from \mbh\ and equation \ref{e:tremaine}, or, alternatively, \mbh\ too large for \sigco.   If \sigco\ is a valid indicator of \sigstar, then these objects seriously violate the local \mbhsigstar\ relationship.  

\begin{figure}[!t]
   \includegraphics[width=\columnwidth]{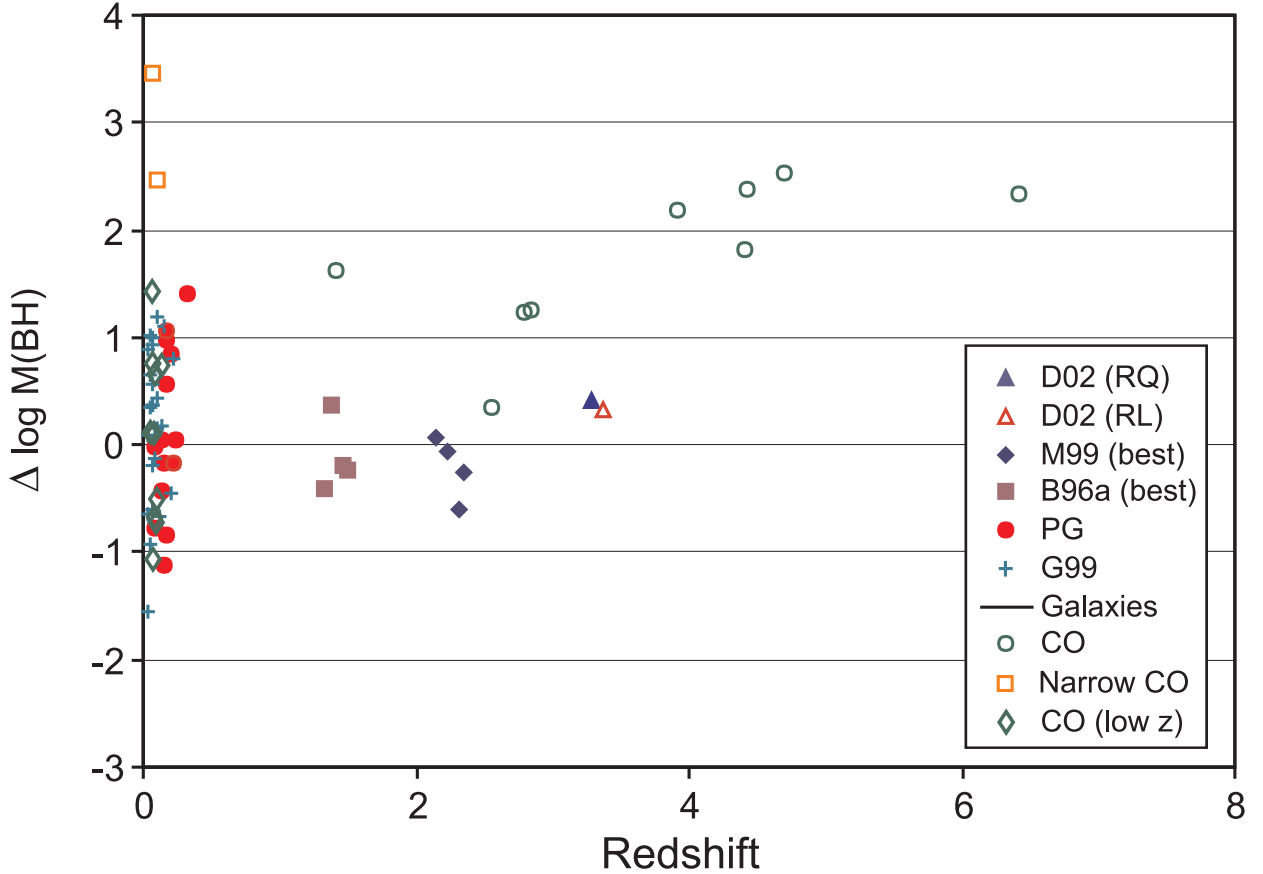}
   \caption{Redshift dependence of $\deltalogmbh \equiv \log\,\mbh - \log\,\mstarsig$ (see text) for the high and low redshift samples. Symbols are as in Figure 3.}
   \label{f:co_dmbh}
\end{figure}

 Figure \ref{f:co_dmbh} shows the redshift dependence of \deltalogmbh.  The highest redshift objects, with $z > 4$, systematically lie nearly two orders of magnitude in \mbh\ above the local \mbhsigstar\ relationship (equation \ref{e:tremaine}, which corresponds to $\deltalogmbh = 0$.)  
 The four CO objects with $z$ between 1.4 and 2.8 (``group 1'') have a mean $\deltalogmbh\ \approx 1.1$, and the five high redshift CO objects  (``group 2'') have $\deltalogmbh \approx 2.2$.  The increase in \deltalogmbh\ from group 1 to group 2 results from an increase in \mbh\  not accompanied by an increase
in \sigco.  These two groups respectively have mean values log~\mbh\ =
8.7, 9.8 and log~\wco\  = 2.16, 2.15.  The two groups have log $L_\mathrm{bol} /L_\mathrm{Ed} $ = -0.5, -0.35, where  we take
$L_\mathrm{bol} \approx 9\lamLlam$ (Kaspi et al. 2000)
and $L_\mathrm{Ed}$ is the Eddington limit.

The radius of the CO emission in high redshift quasars, when known, is typically one or two kpc (Table 3 here and Table 2 of SV05).  How does this compare with the radii at which \sigstar\ is measured in the local galaxies that define the local \mbhsigstar\ relationship?  For the black holes above $10^9~\msun$ included in the \mbhsigstar\ studies (Tremaine et al. 2002), the host galaxies are giant ellipticals with an effective radius $R_e \approx 10^{3.2}$ to $10^{3.9}$~pc (Faber et al. 1997).  For example, M87 has $\mbh = 10^{9.5}~\msun$ and  $R_e = 9$~kpc.  The ``NUKER'' group (e.g., Tremaine et al. 2002)  use a luminosity-weighted \sigstar\ measured within $R_e$, corresponding roughly to $R_e/2$;  Ferrarese \& Merritt (2000) use $R_e/8$.  The velocity dispersion of ellipticals and bulges is generally a weak function of radius, typically $\sigstar \propto R^{-0.04}$ (Jorgensen et al. 1994).  For the larger galaxies defining the local \mbhsigstar\ relationship,  \sigstar\ at $\sim1$~kpc generally varies by less than $\sim10$\% (usually a rise) from $R_e/2$ to $R_e/10$ (Cappellari et al. 2005; Gebhardt et al. 2000;  Bender, Saglia, \& Gerhard 1994).   Luminous elliptical galaxies often have light profiles that flatten into a ``core'' a small radii.  However, the core radii are typically less than $0.1\,R_e$ (Faber et al. 1997), and the volume density of stellar luminosity 
(Gebhardt et al. 1996) only flattens to $\sim r^{-1}$ within the break radius (from $r^{-2}$ at larger radii).  Therefore, the influence of the core on the circular velocity at radii $\sim1$~kpc should be small.  These considerations suggest that the CO gas probes the galactic potential at suitable radii. 

\section{Dynamical and Gas Masses}

CO maps for these ultraluminous objects often show elongated or double structures suggestive of an inclined disk or merger.
Walter et al. (2004) find an elongated structure in SDSS J1148 ($z$ = 6.4), which they interpret in terms of a disk with a radius $r = 2.5$~kpc
seen at an inclination $i \sim 65$~degrees.  Half the CO emission comes from two hot spots separated by $\sim 1.7$~kpc, each containing $\sim 10^{9.7}~\msun$ of molecular gas.  The disk model gives an enclosed mass of $M_\mathrm {dyn} \approx 10^{10.7}$~\msun, which is largely accounted for by the black hole and the molecular gas.    (For a merger model, SV05 quote $M_\mathrm {dyn} \approx 10^{10.2}~\msun$.)   The balance of the measured dynamical mass falls far short of the
bulge mass $\mbulge \approx 700\mbh \approx 10^{12.2}~\msun$ prescribed by the local black hole -- bulge relationship (Kormendy \& Gebhardt 2001).  This discrepancy remains qualitatively true after allowing for a bulge 
radius larger than the CO radius (Walter et al. 2004).  This is consistent with our conclusions using
\sigco\ as a measure of \sigstar\ .
  
Table~3 shows the molecular gas mass for our  CO QSOs given by SV05 along with the dynamical mass quoted by SV05 or estimated by us for
$r_\mathrm{CO} \approx 1$~kpc.  The mass is given by 
$M_\mathrm{dyn}  = (10^{9.37}~\msun) R_\mathrm{kpc} V_{100}^2$,
where $R$ is the disk radius or half the component separation in a merger model, and
$V_{100}$ is \wco\ or half the component separation in units of 100~\kmps\
(SV05).  The resulting values give an average
$M_\mathrm{dyn}\, \mathrm{sin}^2 i \approx10^{10.5}$ \msun, compared with 
$M_\mathrm{gas} \approx10^{10.9}$ and 
$\mbh \approx10^{9.3}$.  Thus, the situation for SDSS J1148 appears to be typical of these QSOs, with the black hole accounting for roughly one-tenth of the dynamical
mass and making a minor contribution to the CO velocities.   The molecular gas accounts for much of $M_\mathrm{dyn}$.   On average, $M_\mathrm{dyn} $ is
a factor $10^{1.6}$ smaller than the bulge mass $\sim700~\mbh$ expected for
the local black hole--bulge relationship.   This conclusion is strongest for the
objects at $z > 3.8$, which have a mean $\mathrm{log}\, M_\mathrm{dyn}/\mbh
= 0.7$, whereas for the lower redshift CO QSOs this value is 2.2.  This is
consistent with our conclusion above that the highest redshift objects have the largest \deltalogmbh.

\section{Discussion}

These results suggest that, at least for the very large black holes involved in the highest redshift QSOs, rapid black hole growth has occurred at $z > 4$ without the formation of a proportionally massive host galaxy.    Taking into account the possible underestimation of \sigco\ with respect to \sigstar\ (see section 2.3), \deltalogmbh\ should be renormalized down by $\sim0.5$.
This gives $\deltalogmbh \approx 0.5$ and 1.5  for groups 1 and 2,  respectively, much larger than any systematic deviations in \mbhsigstar\ relationship for local galaxies.  

Two uncertainties in the interpretation of the CO line widths involve mergers and disk orientation.  The CO profiles often show double peaks that could be interpreted as mergers or the characteristic profile of a rotating disk.  However, for a merger, the individual components will have \wco\ even narrower than the observed profile.  One of the components presumably corresponds to the observed black hole, and \sigco\ for that component alone will be smaller than the \sigco\ we have used, based on the full line profile.  This makes the mismatch of \mbh\ and \sigco\ even worse.  For a single disk model, face-on orientations might be favored for the high redshift QSOs, perhaps to a greater degree than occurs for the PG quasars discussed above.   However, the CO maps are frequently elongated (see discussion of SDSS J1148 above), and the frequent double peaked profiles might not be observed for closely face-on orientations if there is a substantial turbulent component to the orbital motion. Moreover, we see an order-of-magnitude increase in \deltalogmbh\ between groups 1 and 2, despite abundant molecular gas  in both groups.  If we go so far as to renormalize the CO widths to give $\deltalogmbh\  = 0$ for group 1, this still leaves group 2 with $\deltalogmbh \approx 1$.

The CO results here differ from the findings of SO3 for objects in the redshift range $z = 1$ to 3 using \sigthree.  S03
find that, on average, QSOs at  $z = 1$ to 3 obey  equation \ref{e:tremaine}. The black hole masses are similar to those of our high redshift CO QSOs.   The \oiii\ emission comes from the narrow line region (NLR) whose radius in such luminous QSOs should be similar to that of the molecular gas in our CO QSOs.  There is substantial scatter;  Bonning et al. (2005) find an
rms scatter of 0.13 dex in the use of \sigthree\ as a surrogate for \sigstar, corresponding to 0.5 in \deltalogmbh. This seems inadequate to explain the discrepancy between the CO and [O~III] results, since a number of objects are involved in each sample.  Salviander et al. (2005) discuss \sigthree\ and \mbh\ for five radio-quiet QSOs from Sulentic et al. (2004).  Salviander et al. measured  [O~III] widths from the Sulentic et al. spectra, which have better spectral resolution than the high redshift data used by S03.
These objects, with redshifts of 0.8 to 2.4, have a mean \deltalogmbh\ of 0.5,
somewhat larger than in S03.  On the other hand, Bonning et al. (2005) find a mean
\deltalogmbh\ of -0.4 for 6 radio quiet QSOs in the redshift range 2 to 3 from Shemmer et al. (2004) and Netzer et al. (2004).  Overall, these results are consistent with the findings of S03.
The apparent discrepancy between the CO and \oiii\ results underscores the need for a better understanding of both the CO and \oiii\ line widths in the most luminous QSOs.

Our results for high redshift QSOs contrast with the results of Borys et al. (2005), who 
compare the stellar mass in submillimeter galaxies (SMG's) at $z \approx 2$ with the
central black hole masses.  Using population synthesis models to interpret
the infrared luminosity of the SMG's, Borys et al. derive a typical stellar mass 
$M_* \approx 10^{11.4}~\msun$.  The black hole masses are derived from
the X-ray luminosity on the assumption that the bolometric AGN luminosity is close to the Eddington limit (Alexander et al. 2005).  The black hole masses are
$\sim50$ times {\em smaller} than expected for the galaxy mass and the typical ratio of
$\mbh/\mbulge$ for nearby inactive galaxies.  Borys et al. argue that their
small black holes are consistent with a scenario in which a gas-rich merger triggers a massive starburst and builds a large bulge quickly.  An optical QSO is only visible for a brief final phase when the black hole has grown large, supporting 
a large AGN luminosity that dispels the residual gas and reveals the nucleus.  Our CO QSOs  violate the canonical \mbhsigstar\ relationship in the opposite sense of having black holes too large for their host galaxies.

Do the giant black holes observed at high redshift acquire
commensurate host galaxies through later mergers, gas accretion, and star formation?
The local galaxy luminosity function does not afford a sufficient number of commensurately large galaxies to host the largest black holes observed in QSOs (Netzer 2003; Shields \& Gebhardt 2004).   Some of these black holes evidently remain to this day in comparatively modest galaxies.

CO emission lines are currently the only practical means of 
testing the applicability of the \mbhsigstar\ relation to 
galaxies with redshifts larger than $z ~\sim3$.  The present capability for 
such studies is limited.  The number of high-z galaxies with detected CO 
emission is small and not likely to grow dramatically with present 
telescopes.  Furthermore, the known examples are a flux limited sample, 
mostly aided by gravitational lensing.  New facilities such as the Atacama Large Millimeter Array (ALMA) will allow larger, more 
statistically significant and unbiased studies (Carilli 2005).

\acknowledgments

We thank Erin Bonning, Karl Gebhardt, Reinhardt Genzel, Eric Hooper, Eliot Quataert, Sarah Salviander, Marta Volonteri,  Bev Wills, and Marsha Wolf for helpful discussions and communications.  This work was supported by Texas Advanced Research Program grant 003658-0177-2001 and by NSF grant AST-0098594.

\newpage


\begin{table}[htbp!]
\caption{Low Redshift QSOs}
\begin{tabular}{lccccccccc}

&&&&&&&&\\
\hline
\hline
PG & CO reference & $z$ &  CO & [O~III] & $\hbeta$ & log & $\mathrm{log}$ & $\mathrm{\Delta log}$  \\
&&& FWHM\footnotemark[1]  & FWHM  & FWHM & $\nu L\nu $  & \mbh & \mbh	 \\
&&&$\mathrm{(km/s)}$& $\mathrm{(km/s)}$	& $ \mathrm{(km/s)}$ &   & \msun &  CO \\
&&&&&&&\\
\hline								
1404+226 &	Scoville 03&	0.098	& 312	&904	&880	&44.18	&6.71	&-0.71\\
0050+124	 &     Scoville 03&	0.061	& 433	&                &1240	&44.05	&6.95	&-1.05\\
1440+356	 &     Scoville 03&	0.078	& 400	&504	&1450	&44.24	&7.18	&-0.67\\
1119+120	 &     Evans 01 &        0.050        &   260	&413	&1820	&43.92	&7.22	&0.11\\
2130+099 &    	Scoville 03&	0.061	& 325	&	          &2330	&44.27	&7.61	&0.12\\
1415+451	 &      Evans 01	&	0.114	& 50 		&		&2620	&44.21	&7.68	&3.46\\
0838+770 &	Evans 01	&	0.132	& 90 &	533		& 2790	& 44.17 	& 7.71       &2.46\\
0804+761	 &     Scoville 03&	0.100         & 587	&                 &3070	&44.63     &8.02	&-0.5\\
1229+204	 &     Scoville 03&	0.064	& 177	&                 &3360	&44.16	&7.87	&1.44\\
2214+139 & 	Scoville 03&	0.067	& 325	&404	&4550	&44.4	&8.25	&0.76\\
1426+015	 &      Scoville 03&	0.086	& 422	&463	&6820	&44.45	&8.63	&0.68\\
1613+658 & 	Evans 01	&       0.129	& 490	&541	&8450	&44.7	&8.94	&0.73\\
\\							
Average	&			&	0.087	&323&	537&	3282&	44.28&	7.73&	0.57\\
\\								
\hline
\hline
\end{tabular}
\end{table}


\newpage


\begin{table}[htbp!]
\caption{High Redshift QSOs}
\begin{tabular}{lccccccccc}

&&&&&&&&&\\
\hline
\hline
Name & $z$ & FWHM\footnotemark[1]  &$\mathrm{log\, \sigma_{CO}}$ & log & FWHM & Ref & log & log & $\Delta \mathrm{log}$\\ 
& & CO & & $\mathrm{\nu L\nu}$ & UV &	&	$\mbh$ & $L/L_\mathrm{Ed} $ & $ M$ \\ 
\hline
Q0957+561 &	1.412 &	280 &	2.08 &	45.91 &	3850 &1 &	8.86 &	-0.1 &	1.63\\ 
Cloverleaf	   &	2.558 &	400 &	2.23 &	44.86 &	3288 &2 &	8.2 & 	-0.48&	  0.35 \\ 
RX J0911.4 &	2.796 &	350 &	2.17 &	45.21 &	5690 &3 &	8.85 &	-0.78 &	1.23\\  
SMM J04135 &	2.846 &	340 &	2.16 &	45.32 &	5170 &4 &	8.82 &	-0.65 &	1.26\\  
APM 08279 &	3.911 &	480 &	2.31 &	47.49 &	8680	 &5  &  10.36 &	-0.01	 &      2.18\\  
BR 1335 &	4.407 &	420 &	2.25 &	46.47 &	7886 &6 &	9.77 &	-0.44 &	1.83\\  
BR 0952 &	4.434 &	230 &	1.99 &	46.08 &	5624 &6 &	9.27 &	-0.34 &	2.39\\ 
BR 1202  &	4.694 &	350 &	2.17 &	46.7 &	10900 &6 &	10.16 &	-0.61 &	2.54\\  
SDSS J1148 &	6.419 &	250 &	2.03 &	46.17 &	6000 &7 &	9.37 &	-0.36 &	2.34\\  

Average	& &	344 &	2.15	 & 46.02& 6343 &	&	9.3 & 	-0.42 &	1.75\\
\hline
\hline
\end{tabular}
{\footnotesize {${^{1}}$FWHM and \sigco\  in \kmps.}}  Broad ultraviolet emission line is \mgii\ for
SDSS J1148 and \civ\ for other objects. References for broad line width:
(1) Walsh et al. 1979; (2) Monier et al. 1998;` (3) Bade et al. 1997; (4) Knudsen et al. 2003;
(5) Irwin et al. 1998; (6) Storrie-Lombardi et al. 1996; (7) Willott et al. 2003
\end{table}


\vskip20pt


\begin{table}[htbp!]
\caption{High Redshift QSOs}
\begin{tabular}{lcccc}
\\
\hline
\hline
Name & Lense &$ M_\mathrm{gas}$ & Diam.\footnotemark[1]  & log\\
& mag. & $10^{10}\mathrm{M_{\odot}}$ & kpc  & $M_\mathrm{dyn}\,\mathrm{sin}^2\, i$\\	
\hline
Q0957+561 & 3 & 0.4 & &  10.26\\
Cloverleaf	  & 11 & 1.9 &1.5 &  10.45\\
RX J0911.4  & 22 & 0.4 & &  10.46\\
SMM J04135 & 1.3 & 13 & &  10.43\\
APM 08279  & 7 & 1.5 & 2 &  10.73\\
BR 1335  & 1 & 6.4 &  (8.7) & 10.61\\
BR 0952         &	 4 & 0.5 & &  10.09\\
BR 1202  &   3 & 7.9 &  (1.9) &  10.46\\
SDSS J1148  & 2 & 1 &  (1.7) & 10.16 \\
\hline
\hline
\end{tabular}\\
{\footnotesize {${^{1}}$Disk diameter; parenthesis denotes component separation in merger model.}}
\end{table}

\end{document}